\documentclass[conference]{IEEEtran}
\IEEEoverridecommandlockouts
\usepackage{cite}
\usepackage{amsmath,amssymb,amsfonts}
\usepackage{algorithmic}
\usepackage{graphicx}
\usepackage{textcomp}
\usepackage{xcolor}
\usepackage[caption=false,font=footnotesize]{subfig}
\def\BibTeX{{\rm B\kern-.05em{\sc i\kern-.025em b}\kern-.08em
    T\kern-.1667em\lower.7ex\hbox{E}\kern-.125emX}}
\begin{document}

\title{Analytical Characterization of Radial Resolution at Oblique Angles for Near-Field Communications\\

\thanks{This work was supported by the European Union (HORIZON-MSCA-2023-DN-01), Grant Agreement 101169044 - TeraWireless.}
}

\author{\IEEEauthorblockN{1\textsuperscript{st} Viviana Centritto}
\IEEEauthorblockA{\textit{Department of Digital Systems} \\
\textit{University of Piraeus}\\
Piraeus, Greece \\
vcentritto@unipi.gr}
\and
\IEEEauthorblockN{2\textsuperscript{nd} Sotiris Droulias}
\IEEEauthorblockA{\textit{Department of Digital Systems} \\
\textit{University of Piraeus}\\
Piraeus, Greece \\
sdroulias@unipi.gr}
\and
\IEEEauthorblockN{3\textsuperscript{rd} Angeliki Alexiou}
\IEEEauthorblockA{\textit{Department of Digital Systems} \\
\textit{University of Piraeus}\\
Piraeus, Greece \\
alexiou@unipi.gr}
}

\maketitle

\begin{abstract}
Exploiting radial resolvability in near-field communications requires analytical expressions that directly map physical beam parameters to resolution-controllable focal regions. However, beam characterization at oblique angles produces complex algebraic expressions that mask these critical interactions. In this work, we introduce an electromagnetic-aware analytical model built upon a tight upper bound of the spatial power density. This approach yields compact, closed-form solutions for the peak intensity position and radial focal extent, enabling the design of contiguous multi-user focal regions with controlled radial resolution. Performance evaluations confirm that the framework closely tracks exact solutions while facilitating system dimensioning and inter-beam interference management across varying steering angles.
\end{abstract}


\section{Introduction}
\label{sec:introduction}
The transition to Extremely Large Antenna Arrays (ELAAs) at terahertz (THz) frequencies in next-generation architectures extends the conventional far-field boundary, shifting transceiver operations into the radiative near-field \cite{bjornson2019massive, zhang20236g, cui2022near,ye2024extremely}. Within this regime, pronounced spherical wavefront curvature couples the angular and radial domains, transforming directional beamforming into three-dimensional beamfocusing. By concentrating radiated power into distinct focal spots, the system can serve multiple users on the same angular trajectory at different radial distances \cite{liu2024near,liu2025near,zhou2015spherical,ouyang2024primer,zhang2022beam}. However, exploiting this radial resolvability demands the rigorous characterization of focal regions \cite{droulias2024near, bjornson2021primer, manohar2024resolution, rao2024general, centritto2026analytical}. This is further complicated at oblique angles, where a restricted radial span causes focal regions to overlap, resulting in severe inter-beam interference \cite{centritto2026radial}. Multi-user coverage therefore relies on analytical models that translate physical beam parameters into resolution-controllable focal areas. 

Several studies have applied the Fresnel approximation to characterize near-field beamfocusing, deriving analytical expressions for the 3 dB beam depth (BD) that capture its dependence on array geometry, aperture dimensions, and observation angle \cite{bjornson2021primer,kosasih2024finite,hussain2026analyzing}. Further investigations leverage these models to establish scaling laws for the 3 dB BD, maximum finite-depth range, and resolvable spots across varying aperture aspect ratios and SIMO/MIMO processing schemes \cite{wachowiak2025sizing,wachowiak2026analysis}. Despite providing fundamental insights, these frameworks inherently constrain the focal boundary to a static half-power threshold. Consequently, they lack the flexibility required to generate multi-user configurations driven by specific, tunable interference margins.
\begin{figure}[t!]
    \centering
    \includegraphics[width=1.0\columnwidth]{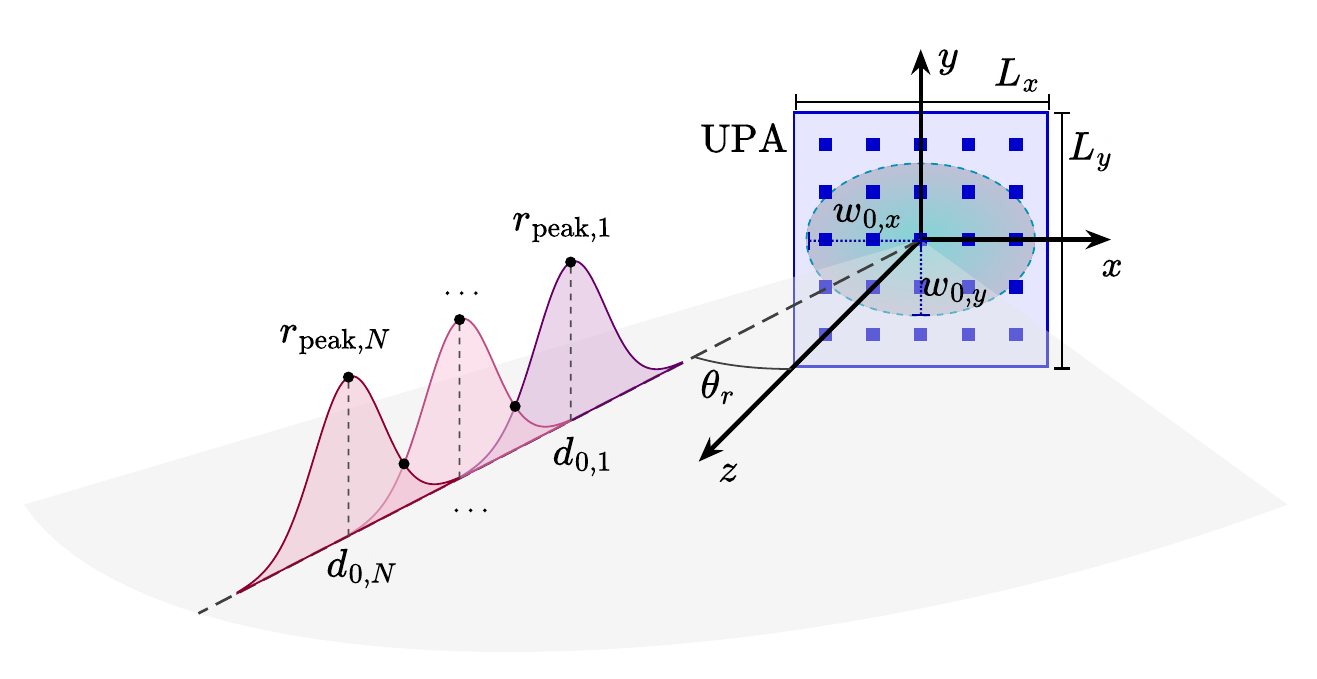}
    \caption{System model of an $L_x \times L_y$ UPA in the $xy$- plane with an elliptical beam footprint ($w_{0,x}, w_{0,y}$), synthesizing $N$ contiguous focal regions at focal distances $d_{0,1}, \cdots, d_{0,N}$ (with peak intensity positions  $r_{\text{peak},1}, \dots, r_{\text{peak},N}$) along the steering angle $\theta_{r}$ in the $xz$- plane.}   
    \label{fig:system_model}
\end{figure}
As an alternative to Fresnel approximations, initial efforts to assess the radial resolution of focused beams using the Angular Spectrum Representation (ASR) were introduced in \cite{droulias2022reconfigurable} for Reconfigurable Intelligent Surfaces (RIS). The authors in \cite{droulias2024near} subsequently extended this ASR framework to evaluate beamfocusing at oblique angles, deriving analytical expressions for the spatial power density and the Full-Width-Half-Maximum (FWHM) metric to quantify axial resolution. More recently, \cite{centritto2026analytical} introduced a tunable mechanism to regulate inter-beam overlap, enabling the synthesis of sequential focal regions with tailored resolution along the broadside axis. Furthermore, closed-form expressions were derived in \cite{centritto2026radial} for the radial and angular boundaries of steered beams to analytically characterize their spatial resolution, allowing iterative numerical methods to generate contiguous focal regions across both domains. Nevertheless, an analytically tractable model that explicitly relates the system's radial resolution to the physical beam parameters at oblique angles has not been derived.

To bridge this gap, this work proposes an electromagnetic-aware analytical model that overcomes the algebraic complexities of characterizing multi-user focal regions at oblique angles. Building upon \cite{centritto2026radial}, we derive a tighter mathematical upper bound on the exact spatial power density, reducing the complex radial boundary formulations into compact, closed-form expressions that are validated against an exact numerical baseline. This tractable bound enables the analytical characterization of adjacent focal regions, granting direct control over radial resolution through the dynamic tuning of beam parameters and inter-beam overlap. Serving as a set of closed-form design guidelines, this framework facilitates the dimensioning of near-field deployments without requiring exhaustive computational sweeps, thereby revealing fundamental trade-offs among physical aperture parameters, user capacity, and achievable spectral efficiency. 

\begin{figure}[!t]
    \centering
    \includegraphics[width=1\columnwidth]{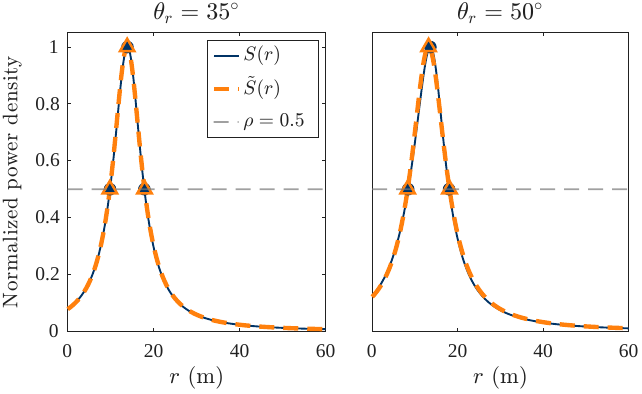}%
    \caption{Normalized power density profiles at $\theta_{r}=35^{\circ}$ and $50^{\circ}$, demonstrating that the proposed model tightly tracks the peak intensity position and the radial boundaries at the inter-beam overlap $\rho = 0.5$.}  
    \label{fig:power_density_comparison}

\end{figure}

\section{System model and analytical framework}
\label{sec:framework}
We consider a uniform planar array (UPA) located at the origin within the $xy$- plane. The array comprises $L_x$ and $L_y$ radiating elements along the $x$- and $y$- axes, respectively, and has a physical size that is significantly larger than the beam's footprint radius. The field amplitude across the array surface follows a Gaussian distribution with an electric field magnitude $E_0$. As illustrated in Fig. \ref{fig:system_model}, the beam exhibits an elliptical footprint on the UPA defined by radii $w_{0,x}$ and $w_{0,y}$. This geometry is governed by a baseline radius $w_0$ and an ellipticity parameter $\theta_i$ that dictates the aspect ratio via $w_{0,y} = w_0$ and $w_{0,x} = w_0/\cos\theta_i$; for the case $\theta_i = 0$, the footprint simplifies to a circle. To perform beamfocusing within the $xz$- plane, the UPA applies a phase profile across its aperture to radiate a converging spherical wavefront. By concentrating the radiated power at a specific focal distance $d_0$ and steering angle $\theta_r$, the system generates a spatially confined focal region dedicated to a user. Following the ASR derivation in \cite{droulias2024near,centritto2026radial}, the power density $S(r)$ along the steering direction ($x = r \sin\theta_r$, $y = 0$, $z = r \cos\theta_r$) is expressed as
\begin{equation} \label{eq:Sr_1}
S(r) = \frac{|E_0|^2}{2\eta_0}\frac{1}{D(r)},
\end{equation}
where $\eta_0$ is the free-wave impedance. The denominator factor $D(r)$ is analytically defined as
\begin{equation} \label{eq:Dr}
   D(r) = \sqrt{\left(1-\frac{r}{d_0}\right)^2+\left(\frac{r}{z_R} \gamma\right)^2} \cdot \sqrt{\left(1-\frac{r}{d_0}\right)^2+\left(\frac{r}{z_R}\right)^2}.
\end{equation}
In this formulation, $z_R = \frac{kw_0^2}{2}$ denotes the Rayleigh range, while the factor $\gamma = \frac{\cos^2\theta_i}{\cos^2\theta_r}$ accounts for the spatial scaling of the beam's footprint based on the ellipticity parameter $\theta_i$ and the steering angle $\theta_r$.

To complement the analytical framework in \cite{centritto2026radial}, which models (\ref{eq:Dr}) via a fourth-degree polynomial that masks the underlying parameter dependencies and requires numerical root-finding for multi-user designs, we seek to define the system's radial resolution as a direct function of the beam parameters $d_0$, $z_R$, and $\theta_r$. By approximating (\ref{eq:Dr}) using the Cauchy-Schwarz inequality ($| \langle \mathbf{x}, \mathbf{y} \rangle| \leq\|\mathbf{x}\| \cdot \|\mathbf{y}\|$), we derive a simplified expression for the radial focal extent. Defining the vectors $\mathbf{x} = (1- \frac{r}{{d_0}}, \frac{r}{{z_R}}\gamma)$ and $\mathbf{y} = (1- \frac{r}{{d_0}}, \frac{r}{{z_R}})$, $D(r)$ can now be expressed as

\begin{equation}
    D(r) =  \|\mathbf{x}\| \cdot \|\mathbf{y}\| \geq  | \langle \mathbf{x}, \mathbf{y} \rangle|.
\end{equation} 
\begin{figure}[!t]
    \includegraphics[width=0.98\columnwidth]{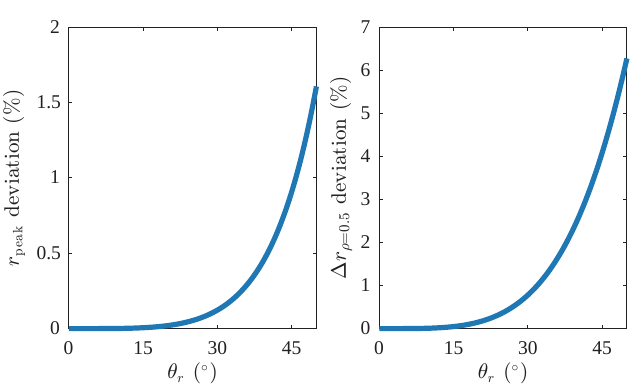}%
    \caption{Relative percentage deviation of the proposed model compared to the exact solution. The evaluation shows the error for the peak intensity position ${r}_{peak}$ and the radial extent $\Delta r_{\rho=0.5}$ as a function of $\theta_r$.} 
    \label{fig:rel_error}
\end{figure}
\begin{figure*}[t!]
    \centering
    \subfloat[]{
        \includegraphics[width=0.45\linewidth]{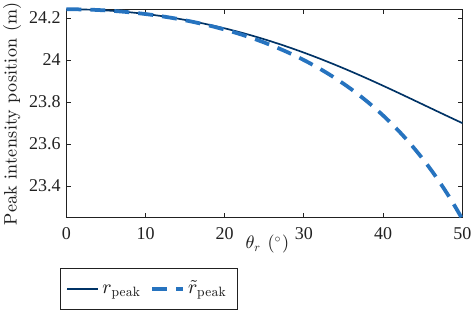}
        \label{fig:r_peak_comparison}
    } 
    \hfill
    \subfloat[]{
        \includegraphics[width=0.45\linewidth]{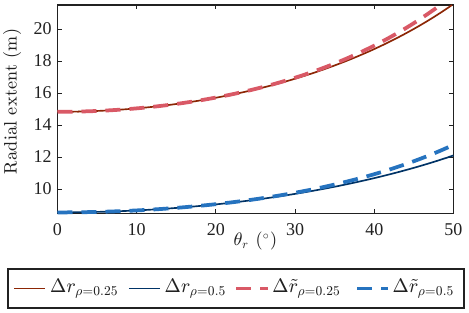}
        \label{fig:delta_r_comparison}
    }   
    \caption{Impact of the steering angle $\theta_r$ on the characterization of an isolated focal region, comparing the exact solution against the approximated model. (a) Behavior of the peak intensity position ($r_{\text{peak}}$ vs. $\tilde{r}_{\text{peak}}$). (b) Radial extent ($\Delta r_{\rho}$ vs. $\Delta \tilde{r}_{\rho}$), illustrating the coarsening of the radial resolution at oblique angles evaluated at $\rho = 0.25$ and $0.5$. Fixed system parameters are $f=150$ GHz, $w_0=0.3$ m, $d_0=25$ m.}
    \label{fig:resolution_comparison}
\end{figure*}
Equality holds when the two vectors are linearly dependent, which occurs where $\gamma = 1$, yielding $\mathbf{x} = \beta \cdot \mathbf{y}$ with $\beta = 1$. Consequently, for scenarios when $\gamma \approx 1$, we can apply the approximation $| \langle \mathbf{x}, \mathbf{y} \rangle| \approx \|\mathbf{x}\| \cdot \|\mathbf{y}\|$. Hereafter, tilde-accented variables denote values directly derived from the proposed analytical approximations, whereas their unmarked counterparts represent the exact numerical solution. Under this convention, expanding $| \langle \mathbf{x}, \mathbf{y} \rangle|$ for $\gamma \approx 1$ yields a tractable upper bound for the power density 
\begin{equation} \label{eq: Sr_Cauchy}
    \tilde{S}(r) = \frac{|E_0|^2}{2\eta_0 | \langle \mathbf{x}, \mathbf{y} \rangle|} = \frac{|E_0|^2}{2\eta_0} \left[ \left( 1-\frac{r}{d_0} \right)^2 + \left( \frac{r}{z_R} \right)^2 \gamma \right]^{-1}.
\end{equation}
By differentiating the denominator in (\ref{eq: Sr_Cauchy}) and solving for $r$, we obtain the peak intensity position  
\begin{equation}\label{eq:r_max_cauchy}
    \tilde{r}_{\text{peak}} = \frac{d_0}{1+(\frac{d_0}{z_R})^2 \gamma}.
\end{equation}
Evaluating $\tilde{S}(r)$ at this position yields the maximum approximated power density
\begin{equation}
    \tilde{S}_{\max} = S(r=\tilde{r}_{\text{peak}}) = \frac{\vert{}E_0\vert{}^2}{2\eta_0} \left( 1 + \frac{1}{\gamma} \left(\frac{z_R}{d_0}\right)^2 \right).
\end{equation}
For the $n$-th focal region, where $n = 1,2,\dots, N$, the radial boundaries are determined by solving $\tilde{S}(r) =\rho \tilde{S}_{\max}$ for a tunable inter-beam overlap parameter $\rho \in (0,1)$, which yields the radial interval
\begin{equation} \label{eq:interval}
    \left[ \tilde{r}_{\text{start},n}, \tilde{r}_{\text{end},n} \right] = \left[ \frac{\frac{z_R^2}{d_0,n} - z_R \sqrt{\gamma} \sqrt{\frac{1-\rho}{\rho}}}{\gamma+\left(\frac{z_R}{d_0,n}\right)^2}, \frac{\frac{z_R^2}{d_0,n} + z_R \sqrt{\gamma} \sqrt{\frac{1-\rho}{\rho}}}{\gamma+\left(\frac{z_R}{d_0,n}\right)^2} \right].
\end{equation}
The difference between these boundaries defines the radial extent, $\Delta \tilde{r}_{\rho,n} = \tilde{r}_{\text{end},n} - \tilde{r}_{\text{start},n}$, which dictates the system's radial resolution
\begin{equation}\label{eq:Delta_r_cauchy}
    \tilde{\Delta r_{\rho,n}} = \frac{2z_{R}\sqrt{\gamma}\sqrt{\frac{1-\rho}{\rho}}}{\gamma+(\frac{z_{R}}{d_0,n})^{2}}.
\end{equation}
%Together, these expressions characterize the focal region explicitly in terms of the focal distance $d_0$, the Rayleigh range $z_R$, and the angular variation $\theta_r$.

The proposed upper bound is validated by comparing $S(r)$ and $\tilde{S}(r)$ profiles across varying steering angles, assuming a circular footprint ($\theta_i = 0$). As illustrated in Fig. \ref{fig:power_density_comparison}, the bound remains tight to the exact numerical evaluation, accurately capturing the peak intensity and radial boundaries of the focused beams at $\theta_r = 35^\circ$ and $50^\circ$. Since the approximation assumes $\gamma \approx 1$, steering further off-axis causes $\gamma$ to deviate from unity, thereby increasing the model's error. This behavior is quantified in Fig. \ref{fig:rel_error}, which demonstrates that the deviations for the peak intensity position ${r}_{_\text{peak}}$ and the radial resolution ${\Delta}{r} _{\rho = 0.5}$ remain negligible at narrow angles. However, these errors increase progressively beyond 45$^\circ$, reaching $0.9$\% and $4.08$\% at $45$$^\circ$. Consequently, to ensure reliable multi-user design, we restrict the operational steering range to approximately $\theta_r$ = $50$$^\circ$ \mbox{($\gamma = 2.42$)} for $\theta_i = 0$.

\section{Synthesis of multiple focal regions}
\label{sec:synthesis}
From (\ref{eq:interval}) and (\ref{eq:Delta_r_cauchy}), $N$ non-uniform resolution focal regions with a fixed radius $w_0$ and varying extents $\Delta\tilde{r}_{\rho}$ are generated at focal distances $d_{0,1},d_{0,2} \dots, d_{0,N}$ ($d_{0,N}$ farthest from the UPA). Generalizing the broadside framework in \cite{centritto2026analytical} to oblique angles, the boundary contiguity condition $\tilde{r}_{\text{end},n-1} = \tilde{r}_{\text{start},n}$ is enforced for $n=2,\dots,N$. Let $P = [(1 - \rho)/\rho]^{1/2}$ be based on $\rho$, and let $B_n = \tilde{r}_{\text{start},n}$ denote the initial boundary. Substituting $\tilde{r}_{\text{end},n-1}$ into the contiguity condition yields 
\begin{equation}
    B_n = \frac{\frac{z_R^2}{d_{0,n-1}} + z_R \sqrt{\gamma} \, P}{\gamma+(\frac{z_R}{d_{0,n-1}})^2},
\end{equation}
which can be rearranged into the quadratic form 
\begin{equation*}
    (B_n \gamma - z_R \sqrt{\gamma} \, P) d_{0,n-1}^2 - z_R^2 d_{0,n-1} + B_n z_R^2 = 0.
\end{equation*}
Solving for the focal distance $d_{0,n-1}$ gives
\begin{equation} \label{eq:recursive_non_uniform}
    d_{0,n-1} = \frac{z_R \left( \sqrt{z_R^2 - 4 B_n^2 \gamma + 4 B_n z_R \sqrt{\gamma} P} - z_R \right)}{2( z_R \sqrt{\gamma} P - B_n \gamma)}.
\end{equation}
For the farthest focal region, the optimal $d_{0,N}$ is derived by maximizing the corresponding $\tilde{r}_{\text{start}, N}$ \cite{centritto2026analytical}. Differentiating the expression for $B_N = \tilde{r}_{\text{start},N}$ with respect to $d_{0,N}$ yields
\begin{equation} \label{eq:d_N}
    d_{0,N} = \frac{z_R \left(1 - \sqrt{1-\rho}\right)}{\sqrt{\gamma \rho}}.
\end{equation}
\begin{figure*}[!t]
    \centering 
    \subfloat[]{
        \includegraphics[width=0.47\linewidth]{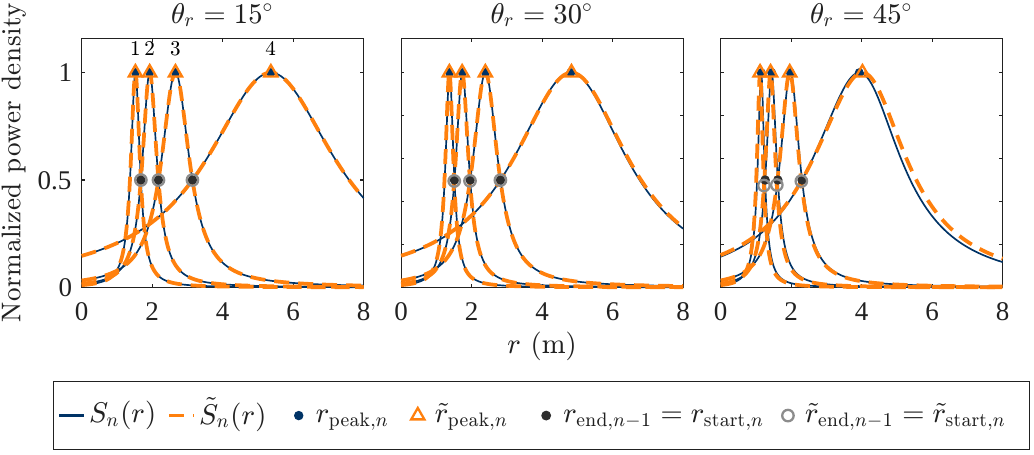}
        \label{fig:non_uniform-chain}
    }
    \hfil 
    \subfloat[]{
        \includegraphics[width=0.48\linewidth]{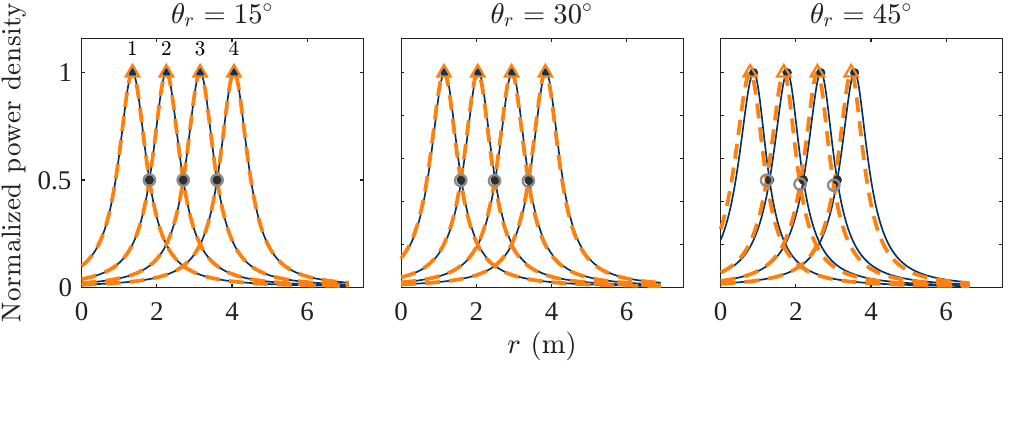}
        \label{fig:uniform-chain}}
    \caption{Resolving in the radial domain with $N = 4$ contiguous focal regions for $\theta_i = 0^\circ$. The exact solution is compared against the proposed approximation across $\theta_r = 15^\circ, 30^\circ$, and $45^\circ$, with an inter-beam overlap of $\rho = 0.5$. Profiles of the synthesized focal regions using (a) non-uniform resolution with a fixed beam radius $w_0$, and (b) uniform resolution with a tunable beam radius $w_{0,n}$.}
\end{figure*}
\begin{figure*}
    \subfloat[]{
        \includegraphics[width=0.473\linewidth]{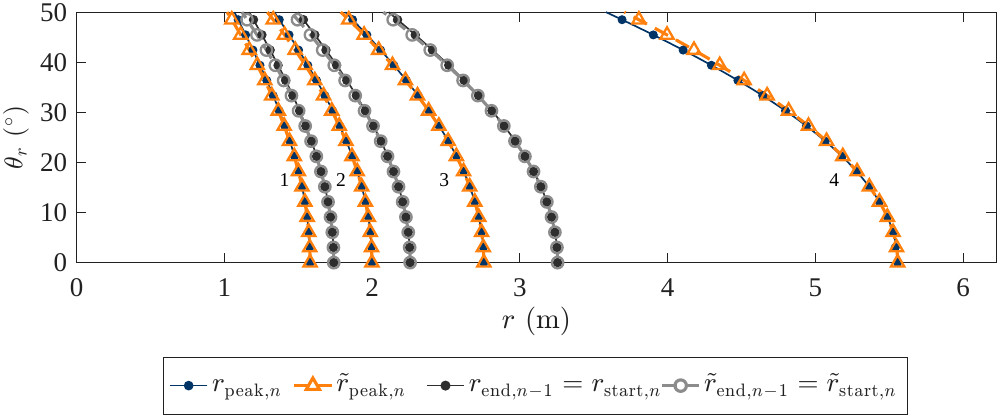}
        \label{fig:non_uniform-chain-r-vs-theta}
    }
    \hfil 
    \subfloat[]{
        \includegraphics[width=0.48\linewidth]{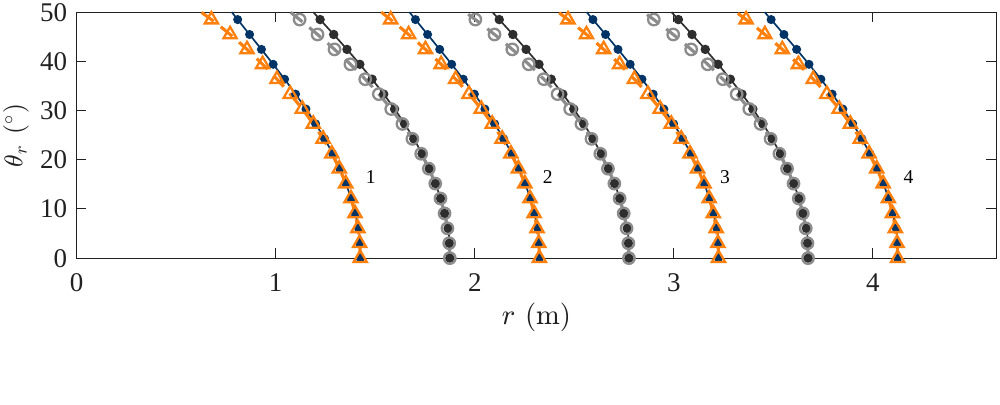}
        \label{fig:uniform-chain-r-vs-theta}
    }
    \caption{Influence of angular variation $\theta_r$ on peak intensity positions $r_{peak}$ and boundary intersection points $r_{\text{end}, n-1} = r_{\text{start}, n}$. (a) Non-uniform resolution. (b) Uniform resolution.}
    \label{fig:beam_synthesis}
\end{figure*}

Analytical expressions for synthesizing focal regions with uniform radial resolution at oblique angles are also of practical interest. A sequence of $N$ focal regions with a constant $w_{\text{FWHM}}$ (i.e., $\Delta \tilde{r}_{\rho=0.5}$, as $\Delta \tilde{r}_{\rho,n} = P w_{\text{FWHM},n}$) intersecting at inter-beam overlap $\rho$ is synthesized by adjusting each beam radius $w_{0,n}$ via its Rayleigh range $z_{R,n}$ \cite{centritto2026analytical}. After fixing the outermost focal distance $d_{0,N}$, computing $z_{R,N}$ from (\ref{eq:Delta_r_cauchy}), and setting $B_N = \tilde{r}_{\text{start},N}$, the recursion in (\ref{eq:recursive_non_uniform}) is adapted to jointly determine the remaining $d_{0,n-1}$ and $z_{R,n-1}$. Combining (\ref{eq:Delta_r_cauchy}) with $\tilde{r}_{\text{end},n-1} = \tilde{r}_{\text{start},n}$ yields
\begin{equation}\label{eq:uniform-d0n-1}
    d_{0, n-1} = \frac{\left(B_n - \frac{\tilde{\Delta r_{\rho}}}{2}\right)^2 + \left(\frac{\tilde{\Delta r_{\rho}}}{2P}\right)^2}{B_n - \frac{\tilde{\Delta r_{\rho}}}{2}},
\end{equation}
\begin{equation}\label{eq:uniform-zRn-1}
    z_{R, n-1} = \sqrt{\gamma} \frac{\left(B_n - \frac{\tilde{\Delta r_{\rho}}}{2}\right)^2 + \left(\frac{\tilde{\Delta r_{\rho}}}{2P}\right)^2}{\frac{\tilde{\Delta r_{\rho}}}{2P}}.
\end{equation}
from which $z_{R,n-1}$ maps directly to the required beam radius via $w_{0,n-1} = \sqrt{2z_{R,n-1}/k}$. Under the finite-aperture constraint, $w_{0,n-1}$ is bounded by $w_{0,\max}$ (with corresponding $z_{R,\max}$) \cite{centritto2026analytical}. Substituting $z_{R,N} = z_{R,\max}$ into (\ref{eq:Delta_r_cauchy}) and solving for $d_{0,N}$ yields the farthest aperture-constrained focal distance
\begin{equation} \label{d_0_N_uniform}
    d_{0,N} = z_{R,\max} \sqrt{\frac{\Delta \tilde{r}_\rho}{2P \sqrt{\gamma} z_{R,\max} - \gamma \Delta \tilde{r}_\rho}}.
\end{equation}
In both synthesis designs, the recursive process terminates at the ASR-analytical validity limit $d_{0,n-1} \ge d_{0,\min}$ derived in \cite{centritto2026radial}, which remains constant for the non-uniform system configuration but varies for each focal region in the uniform case.

\section{Results}
\label{sec:results}
\begin{figure*} [t!]
    \centering
    \subfloat[]{
        \includegraphics[width=0.9\linewidth]{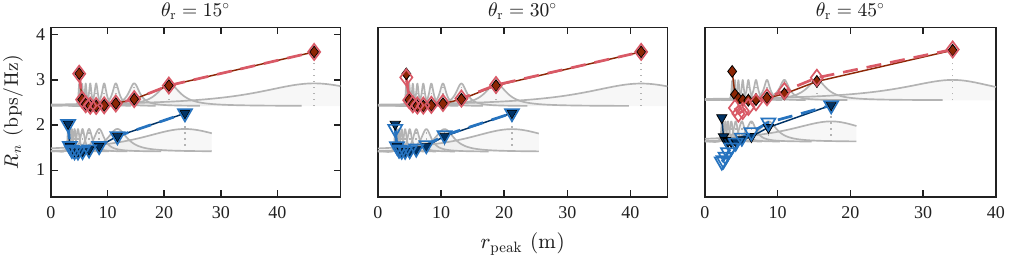}
        \label{fig:per_user}
    }
    
    \subfloat[]{
        \includegraphics[width=0.9\linewidth]{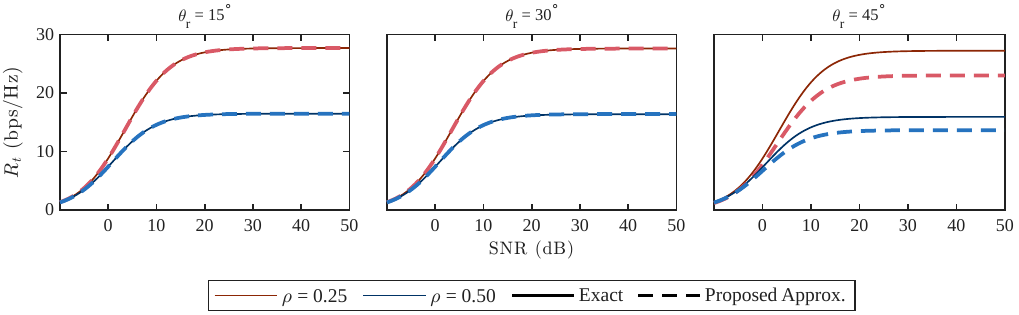}
        \label{fig:system_eval}
    }
    \caption{Performance assessment of a non-uniform radial resolution system supporting a fixed capacity of $N=10$ users. The exact solution is compared against the proposed approximation across steering angles of $\theta_r=15^\circ, 30^\circ$, and $45^\circ$ and inter-beam overlap thresholds of $\rho=0.25$ and $0.50$. (a) Per-user spectral efficiency $R_n$ evaluated across the radial domain $r$ for a fixed SNR of $20$ dB. (b) Sum rate $R_t$ as a function of the SNR. The system configuration operates at $f=150$ GHz with $\theta_i=0$.}
    \label{fig:combined_evaluation}
\end{figure*}

The proposed analytical model yields compact solutions for the peak intensity position $\tilde{r}_{\text{peak}}$ and the radial focal extent $\Delta \tilde{r}_{\rho}$, revealing the critical interplay between the physical beam parameters and the system's radial resolution. Throughout the following examples, a circular footprint ($\theta_i = 0^\circ$) is assumed across all configurations, which reduces $\gamma$ to $\cos^{-2}\theta_r$. To isolate the effects of the steering angle on a single focal region, Fig. \ref{fig:resolution_comparison} depicts an individual beam aimed at a fixed focal distance $d_0 = 25$ m. Driven by the $\cos^{-2}\theta_r$ term in (\ref{eq:r_max_cauchy}), $\tilde{r}_{\text{peak}}$ decreases monotonically with $\theta_r$, shifting the peak intensity position approximately $0.3$ m closer to the array from $\theta_r$ = $0^\circ$ to $30^\circ$, as shown in Fig. \ref{fig:r_peak_comparison}. Furthermore, Fig. \ref{fig:delta_r_comparison} illustrates the impact of the steering angle on the radial extent for a given inter-beam overlap $\rho$. Because $d_0$ is held constant, increasing $\theta_r$ enlarges $\gamma$ in (\ref{eq:Delta_r_cauchy}), which broadens the focal region's radial span and coarsens the radial resolution. For a half-power threshold ($\rho = 0.5$), the focal extent expands by approximately $3$ m as the propagation axis changes from $0^\circ$ to $45^\circ$, while a stricter overlap of $\rho = 0.25$ yields a wider focal extent that increases from $14.8$ m at broadside to $20.37$~m at $\theta_r = 45^\circ$. 

With the isolated, fixed-$d_0$ beam characterized, the analysis focuses on synthesizing $N=4$ contiguous focal regions. Fig. \ref{fig:non_uniform-chain} depicts the non-uniform resolution synthesis using a constant beam radius of $w_0 = 0.1$ m, where the focal distances $d_{0,n-1}$ are computed recursively via (\ref{eq:recursive_non_uniform}). This fixed $w_0$ locks the Rayleigh range $z_R$; therefore, the progressively larger $d_{0,n}$ values for outer users reduce the $(z_R/d_{0,n})^2$ term in (\ref{eq:Delta_r_cauchy}), causing the radial extent $\Delta \tilde{r}_{\rho,n}$ to expand with distance and degrade the radial resolution. To ensure equal resolution across all regions independent of the steering angle, the uniform radial resolution synthesis illustrated in Fig. \ref{fig:uniform-chain} maintains a constant $\Delta \tilde{r}_{\rho =0.5} = 0.9$ m. This is achieved by jointly solving for $d_{0,n-1}$ and $z_{R,n-1}$ via (\ref{eq:uniform-d0n-1}) and (\ref{eq:uniform-zRn-1}), dynamically enlarging the beam radius $w_{0,n}$ with increasing distance. Moreover, Fig. \ref{fig:beam_synthesis} depicts the influence of $\theta_r$ on $\tilde{r}_{\text{peak}}$ and the boundary condition $\tilde{r}_{\text{end,n-1}} = \tilde{r}_{\text{start,n}}$ across the entire beam sequence, where $d_{0,n}$ is shown to be sensitive to the angular variation. In the non-uniform case (Fig. \ref{fig:non_uniform-chain-r-vs-theta}), the outermost focal distance scales as $d_{0,N} \propto \gamma^{-1/2} = \cos\theta_r$ in (\ref{eq:d_N}), moving all $d_{0,n}$ closer to the array at increasingly oblique angles. According to (\ref{eq:r_max_cauchy}), this reduction in $d_{0,n}$ forces the peak intensity positions to displace inward at the same rate; for instance, $r_{\text{peak},4}$ decreases from $5.5$ m at $0^\circ$ to $4$ m at $45^\circ$. Regarding the radial extent $\Delta\tilde{r}_{\rho,n}$ in (\ref{eq:Delta_r_cauchy}), the smaller $d_{0,n}$ values quadratically enlarge the denominator's $(z_R/d_{0,n})^2$ term enough to outweigh the numerator's $\sqrt{\gamma} = \cos^{-1}\theta_r$ growth, thereby causing an increased steering angle to narrow each radial extent and restrict the overall synthesizable radial span. Under the uniform scenario (Fig. \ref{fig:uniform-chain-r-vs-theta}), $d_{0,N}$ is initialized via (\ref{d_0_N_uniform}), which exhibits a non-linear dependence on $\theta_r$ rather than a direct  $\cos\theta_r$ scaling, resulting in a more moderate peak inward shift compared to the non-uniform case, with $r_{\text{peak},4}$ changing from $4.1$ m at $0^\circ$ to $3.4$ m at $45^\circ$. Across both designs, the proposed approximation exhibits agreement with the exact solution, with minor divergence only appearing beyond $\theta_r > 40^\circ$.

\begin{figure} [t!]
    \centering
    \includegraphics[width=0.94\linewidth]{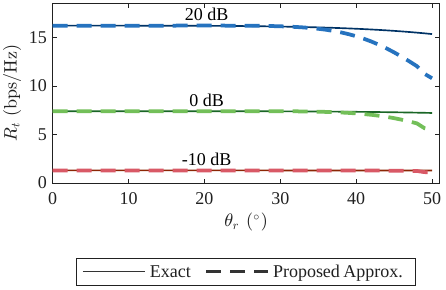}
    \caption{Sum rate $R_t$ analyzed across steering angles $\theta_r$. The plot compares the performance of the exact solution against the proposed approximation across varying SNR scenarios with fixed $\rho = 0.5$.}
    \label{fig:rt_vs_theta}
\end{figure}

To quantify system performance in multi-user configurations, we evaluate the per-user spectral efficiency \mbox{$R_n = \log_2(1 + \text{SINR}_n)$} and the sum rate $R_t = \sum{R_n}$, comparing the proposed analytical model against the exact baseline. Expressed as a function of the signal-to-noise ratio (SNR), the signal-to-interference-plus-noise ratio (SINR) for the $n$-th user is $\text{SINR}_n = \frac{\text{SNR}}{\alpha_n \cdot \text{SNR} + 1}$. Here, the factor $\alpha_n$ is proportional to the interference level experienced by each beam, which is dictated by the inter-beam overlap threshold $\rho$ \cite{centritto2026analytical}. Two non-uniform resolution systems supporting a fixed capacity of $N=10$ users with \mbox{$\rho = 0.25$} ($w_0 = 0.35$ m) and $\rho = 0.50$ ($w_0 = 0.21$ m) are considered. At \mbox{SNR = 20 dB}, Fig. \ref{fig:per_user} shows the per-user rate evaluated at the true user positions $r_{\text{peak}}$ along the propagation axis $r$. The inner-middle beams ($n=2\text{--}4$) exhibit the lowest spectral efficiency ($\approx$~$2.4$~bps/Hz for $\rho$ = $0.25$ and $\approx$~1.4 bps/Hz for \mbox{$\rho$~=~$0.50$}) due to tightly packed focal regions, which increases inter-beam interference. Conversely, the outer regions become more spaced out with increasing distance, mitigating tail interference and enabling the outermost user to achieve the maximum rate ($\approx$ $3.6$ bps/Hz for $\rho$ = $0.25$ and $\approx$ $2.25$ bps/Hz for $\rho$ = $0.50$). Adopting a stricter isolation improves performance across all users, although at the cost of a larger beam radius and broader radial coverage span.  The sum rate as a function of the SNR is depicted in Fig. \ref{fig:system_eval}, revealing two distinct performance regimes. In the low-SNR regime (SNR $<$ 0 dB), both overlap thresholds yield similar performance. At high SNR (SNR $>$ 20 dB), system capacity becomes interference-limited rather than noise-limited; $\rho = 0.5$ saturates at $16.5$~bps/Hz, whereas stricter isolation reduces interference and increases the achievable sum rate to $27.6$ bps/Hz for $\theta_r<30^\circ$. Additionally, the sum rate across steering angles for $\rho$ = $0.5$ is illustrated in Fig. \ref{fig:rt_vs_theta}, which indicates a more pronounced capacity degradation at highly oblique angles. This degradation occurs because the outermost focal distance $d_{0,N}$ contracts inward with $\theta_r$, compressing the focal regions into a narrower radial span and elevating the adjacent-beam interference. Crucially, across both metrics, the analytical approximation remains indistinguishable from the exact solution up to $\theta_r \approx 35^\circ$, with rate underestimation emerging only at steeper angles $\theta_r > 45^\circ$ in high-SNR regimes.
%Expressed as a function of the signal-to-noise ratio (SNR), the signal-to-interference-plus-noise ratio (SINR) for the $n$-th user is $\text{SINR}_n = \frac{\text{SNR}}{\alpha_n \cdot \text{SNR} + 1}$. Here, $\alpha_n = \frac{\sum_{l \neq n} P_l}{P_n}$ quantifies inter-beam interference, where $P_l$ and $P_n$ are the unnormalized spatial power densities $S(r)$ evaluated at user $n$'s exact peak position, governed by the overlap $\rho$ \cite{centritto2026analytical}.
%two-sided inter-beam interference.

\section{Conclusion}
\label{sec:conclu}
In this work, we developed an electromagnetic-aware analytical model for characterizing near-field radial resolution at oblique angles. By deriving a tight upper bound on the spatial power density, the proposed approach yielded compact closed-form solutions that directly associate radial resolution with beam parameters $w_0$, $z_R$, $d_0$, $\theta_r$, and inter-beam overlap $\rho$. The ability to analytically manipulate these parameter interactions enabled the recursive synthesis of contiguous non-uniform and uniform resolution multi-user focal regions without relying on iterative numerical methods. Essentially, these results validated the framework as an effective tool for evaluating key near-field metrics, regulating inter-beam interference, and executing rapid system dimensioning in complex multi-user environments.

\bibliographystyle{IEEEtran}
\bibliography{bib}

\end{document}